\documentclass[10pt,aps,twocolumn,prc,floatfix,preprintnumbers,superscriptaddress,notitlepage,nofootinbib]{revtex4-1}
\usepackage{amssymb, amsmath, bm}

\usepackage[english]{babel}
\usepackage[utf8]{inputenc}
\usepackage{blindtext}
\usepackage{graphicx}
\usepackage{hyperref}
\hypersetup{
    colorlinks,
    linkcolor={red!50!black},
    citecolor={blue!50!black},
    urlcolor={blue!80!black}
}
\usepackage{physics}
\usepackage{xcolor}
\usepackage[normalem]{ulem}
\usepackage{float}
\usepackage{comment}

\usepackage{tikz}

\begin{document}

\title{Tomography of Atomic Nuclei}
\author{Noemi Rocco}
\email{rocco@csic.uv.es}
\affiliation{Instituto de Física Corpuscular (IFIC), Consejo Superior de Investigaciones Científicas (CSIC) and Universidad de Valencia
E-46980 Paterna, Valencia, Spain}
\affiliation{Theoretical Physics Department Fermi National Accelerator Laboratory P.O. Box 500 Batavia Illinois 60510 USA}

\author{Carlo Graziani}
\email{carlo@uchicago.edu}
\affiliation{ Mathematics and Computer Science Division, Argonne National Laboratory, Argonne, Illinois 60439, USA}

\author{Alessandro Lovato}
\email{lovato@anl.gov}
\affiliation{Physics Division, Argonne National Laboratory, Argonne, Illinois 60439, USA}
\affiliation{Computational Science Division, Argonne National Laboratory, Argonne, Illinois 60439, USA}
\affiliation{INFN-TIFPA Trento Institute for Fundamental Physics and Applications, Trento, Italy}
\affiliation{Instituto de Física Corpuscular (IFIC), Consejo Superior de Investigaciones Científicas (CSIC) and Universidad de Valencia
E-46980 Paterna, Valencia, Spain}

\author{Anthony J. Tropiano}
\email{atropiano@anl.gov}
\affiliation{Physics Division, Argonne National Laboratory, Argonne, Illinois 60439, USA}

\author{R.~B. Wiringa}
\email{wiringa@anl.gov}
\affiliation{Physics Division, Argonne National Laboratory, Argonne, Illinois 60439, USA}

\date{\today}

\begin{abstract}
We carry out continuum quantum Monte Carlo calculations of the quantum-mechanical Wigner distribution functions of selected nuclei, up to $^{16}$O. These distributions provide a form of quantum tomography of the spatial and momentum structure of the system. They also help identify the location of high-momentum regions in atomic nuclei and provide insight into the onset of alpha clustering. Besides their intrinsic interest, these distributions will be useful for neutrino event generators, as they correlate the positions and momenta of nucleons in the initial target state. To facilitate their application, we address the need to store them compactly by developing an accurate Gaussian Process emulator that automatically preserves their normalization.
\end{abstract}

\maketitle

\section{Introduction}
Nuclear continuum quantum Monte Carlo (QMC) methods provide an accurate multiscale description of atomic nuclei, from collective excitations and alpha clustering to short-range correlations, in terms of the underlying interactions among constituent protons and neutrons~\cite{Carlson:2014vla,Gandolfi:2020pbj}. In addition to integrated observables, such as energies, radii, and transition amplitudes, these methods also provide direct access to probability distributions in coordinate and momentum space~\cite{Wiringa:2013ala, Piarulli:2022ulk}. An even more complete characterization of nuclear structure, however, is provided by joint position-momentum, or phase-space, distributions, which offer a tomographic representation of the nucleus.

In classical mechanics, the state of a particle is completely specified by its position and momentum. For example, the single-particle properties of a classical gas can be described by a phase-space probability distribution whose time evolution is governed by the Boltzmann equation~\cite{Ji:2003ak}. The first quantum-mechanical phase-space distribution was introduced by Wigner in 1932~\cite{Wigner:1932eb}. Its equation of motion reduces to the classical one in the limit $\hbar \to 0$~\cite{Nikolakopoulos:2025fgw}. Because the uncertainty principle forbids the simultaneous determination of a particle's position and momentum, the Wigner distribution is a quasi-probability distribution and is therefore not positive definite.

Wigner distributions provide a unique perspective on the spatial and momentum structure of quantum systems and have found wide-ranging applications, from quantum information science and quantum electronics, where they are used to model electron transport, to quantum chemistry, where they are employed to calculate the static and dynamical properties of many-body quantum systems, and to signal processing, where they are used to study wave propagation through complex media~\cite{Weinbub:2018sjs}. The study of Wigner distributions has recently attracted considerable attention in nonperturbative quantum chromodynamics (QCD)~\cite{Belitsky:2003nz,Lorce:2011kd,Lorce:2011ni,Hagiwara:2017fye}. In this context, the QCD Wigner distribution encodes the joint position-momentum distributions of partons in the nucleon and is often regarded as the ``mother'' distribution, since several other partonic distributions can be obtained from it through suitable projections or reductions~\cite{Lorce:2011kd}.

To the best of our knowledge, this work presents the first \textit{ab initio} calculations of Wigner distributions for selected nuclei, up to $^{16}$O. Previous calculations of nuclear Wigner distributions were either limited to the deuteron~\cite{Neff:2016ajx} or performed within mean-field approaches~\cite{Prakash:1981ilg,Bonasera:1993zz,Nikolakopoulos:2025fgw} or within the lowest-order correlation operator approximation to account for short-range correlations~\cite{Cosyn:2021ber}. In contrast, quantum Monte Carlo methods treat long- and short-range correlations on the same footing and are therefore well suited to describe the rich structural features of nuclei, including clusterization in light systems. These features are revealed by the corresponding phase-space distributions.

Our calculations are based on the highly accurate phenomenological Argonne $v_{18}$~\cite{Wiringa:1994wb} nucleon-nucleon (NN) interaction and its reprojection, Argonne $v_8^\prime$~\cite{Wiringa:2002ja}, supplemented by either the Urbana IX or Urbana X three-nucleon (3N) force~\cite{Pudliner:1995wk,Wiringa:2013ala}. These Hamiltonians are employed in variational Monte Carlo (VMC)~\cite{Lomnitz-Adler:1981dmh} and auxiliary-field diffusion Monte Carlo (AFDMC)~\cite{Schmidt:1999lik} methods, which provide highly accurate solutions of the quantum many-body problem. For $^4$He, results obtained with the VMC and AFDMC approaches are compared directly.

In addition to their intrinsic interest, Wigner distributions can be readily incorporated into neutrino event generators, in particular {\sc Achilles}~\cite{Isaacson:2020wlx,Isaacson:2022cwh}. In the current implementation, the spatial distributions of neutrons and protons are sampled from nuclear configurations obtained from QMC calculations, fully retaining correlation effects. Their initial momenta, however, are sampled independently of position, thus neglecting spatial-momentum correlations. Employing Wigner distributions would remove this approximation. Particularly important in this context is the use of Gaussian processes (GPs)~\cite{Rasmussen:2006} to store and interpolate these distributions efficiently and compactly. As a specific development within the GP framework, we also address the problem of imposing an integral normalization constraint, which, to the best of our knowledge, has not been considered previously in the literature.

The present article is organized as follows. In Sec.~\ref{sec:hamiltonians}, we summarize the nuclear Hamiltonians employed in the calculation of the Wigner distributions, while Sec.~\ref{sec:qmc} briefly reviews the VMC and AFDMC methods. In Sec.~\ref{sec:wigner}, we introduce the Wigner distributions, describe how they are computed within QMC, and discuss their relation to the spatial and momentum distributions. Section~\ref{sec:GPs} describes the Gaussian-process procedure used to store and interpolate the Wigner distributions reliably. In Sec.~\ref{sec:results}, we present and discuss results for selected nuclei. Finally, Sec.~\ref{sec:conclusions} contains our conclusions and an outlook for future work.

\section{Nuclear Hamiltonians}
\label{sec:hamiltonians}
The interactions among point-like nucleons are modeled by the non-relativistic Hamiltonian\footnote{We use natural units in which $\hbar=c=1$.}
\begin{align}
H = \sum_i^A \frac{{\bf p}_i^2}{2m} + \sum_{i<j}^{A} v_{ij}
 + \sum_{i<j<k}^A V_{ijk} \ ,
\label{H:A}
\end{align}
where ${\bf p}_i$ and $m$ denote the momentum of the $i$-th nucleon and its
mass, and the potentials $v_{ij}$ and $V_{ijk}$ describe nucleon-nucleon ($NN$) and three-nucleon ($3N$) interactions, respectively. Highly-realistic, phenomenological $NN$ potentials, such as the Argonne $v_{18}$ (AV18)~\cite{Wiringa:1994wb}, are fitted to reproduce the observed properties of the
two-nucleon system, including the deuteron binding energy, magnetic moment and electric quadrupole moment, as well as the data obtained from the measured NN scattering cross sections\textemdash and reduce to Yukawa's one-pion-exchange potential at large distance. They are defined in coordinate space as
\begin{equation}
v_{ij} = v^\gamma_{ij} + \sum_{p=1}^{18} v^p(r_{ij})O^p_{ij}
\label{eq:vNN}
\end{equation}
with $r_{ij}= |{\bf r}_i - {\bf r}_j|$.  Here $v^\gamma$ denotes a complete electromangetic interaction including Coulomb, magnetic moment, and vacuum polarization terms.
The bulk of the strong $N\!N$ interaction is encoded in the first eight operators
\begin{equation}
O^{p=1-8}_{ij}= [1, \sigma_{ij},S_{ij},\mathbf{L}\cdot\mathbf{S}]\otimes [1,\tau_{ij} ] \, ,
\label{eq:oper}
\end{equation}%
where we introduced $\sigma_{ij}={\bm \sigma_i}\cdot{\bm \sigma_j}$ and $\tau_{ij}={\bm \tau_i}\cdot{\bm \tau_j}$ with ${\bm \sigma}_i$ and ${\bm \tau}_i$ being the Pauli matrices acting in the spin and isospin space. The tensor operator is given by
\begin{equation}
S_{ij}= \frac{3}{r^2_{ij}}({\bm \sigma}_i\cdot {\bf r}_{ij})({\bm \sigma}_j \cdot{\bf r}_{ij})- \sigma_{ij}\, ,
\end{equation}
while the spin-orbit contribution is expressed in terms of the relative angular momentum $\mathbf{L}=\frac{1}{2 i} (\mathbf{r}_i -\mathbf{r}_j) \times (\boldsymbol{\nabla}_i - \boldsymbol{\nabla}_j)$ and the total spin $\mathbf{S}=\frac{1}{2}({\bm \sigma}_i+{\bm \sigma}_j)$ of the pair. AV18 contains six additional charge-independent operators, corresponding to $p=9\text{--}14$, involving $L^2$ and $(\mathbf{L}\cdot\mathbf{S})^2$, while the operators with $p=15\text{--}18$ account for charge dependence and charge asymmetry.

Part of the results presented in this work were obtained using a simplified version of the full AV18 potential, dubbed Argonne $v^\prime_8$ (AV8P). AV8P contains only the first eight operators of Eq.~\eqref{eq:oper} and is a reprojection, rather than a simple truncation, of AV18~\cite{Pudliner:1997ck}. The AV8P potential correctly reproduces all $S$- and $P$-wave phase shifts and preserves the deuteron; discrepancies with AV18 appear only in higher partial waves. When employing the AV8P potential, we assume the electromagnetic component of the NN potential to only include the Coulomb force between finite-size protons. 

As is common practice when using $NN$ interactions of the Argonne family, we employ the Urbana IX (UIX)~\cite{Pudliner:1995wk} or Urbana X (UX)~\cite{Wiringa:2013ala} $3N$ force. The Urbana models include a Fujita--Miyazawa two-pion-exchange P-wave contribution~\cite{Fujita:1957zz}, associated with intermediate virtual $\Delta(1232)$ excitation, together with a phenomenological short-range repulsive term. The two-pion-exchange part is written as the sum of ``anticommutator'' and ``commutator'' terms, given by
\begin{align}
V_{ijk}^{\Delta,a} &= \sum_{\rm cyc} A^{PW}_{2\pi}\{X_{ij}^\pi,X_{jk}^\pi\}\{\tau_{ij},\tau_{jk}\} \, , \nonumber\\
V_{ijk}^{\Delta,c} &= \sum_{\rm cyc} C^{PW}_{2\pi}[X_{ij}^\pi,X_{jk}^\pi][\tau_{ij},\tau_{jk}] \, ,
\label{eq:FM}
\end{align}
where $\sum_{\rm cyc}$ denotes a sum over the three cyclic exchanges of nucleons $i,j,k$.  The operator ${X}^\pi_{ij}$ is defined as 
\begin{equation}
X_{ij}^\pi = T(r_{ij})\, S_{ij} + Y(r_{ij})\sigma_{ij} \, ,
\end{equation}
where the normal Yukawa and tensor functions are consistent with those used in the $NN$ force
\begin{align}
Y(r)&=\frac{e^{-{m_{\pi}}r}}{{m_{\pi}}r} \,\xi(r) \nonumber\\
T(r)&=\left( 1 + \frac{3}{m_{\pi}\,r} + \frac{3}{m_{\pi}^2\, r^2} \right)  Y(r) {\,\xi(r)} \, .
\end{align}
In the latter equation, the short-range regulator is defined as
$\xi(r) = 1 - \exp(-c r^2)$, with $c = 2.1~\mathrm{fm}^{-2}$. In the Urbana models, it is assumed that $C^{PW}_{2\pi} = A^{PW}_{2\pi} / 4$ as in the original Fujita-Miyazawa formulation. In order to correctly reproduce the isospin-symmetric nucleonic matter saturation density~\cite{Lagaris:1981mn}, a repulsive 3N force should also be included in the potential; we use the form
\begin{equation}
V^{R}_{ijk} = A_R \sum_{\rm cyc} T^2(r_{ij}) T^2(r_{jk}) \, .
\end{equation}

The constants $A^{PW}_{2\pi}$ and $A_R$ are determined by fitting the binding energy of $^3$H {and $^4$He} and the saturation density of isospin-symmetric nucleonic matter $\rho_0 = 0.16$ fm$^{-3}$.
The UX model has an additional small two-pion S-wave term
\begin{equation}
V_{ijk}^{SW} = \sum_{\rm cyc} A^{SW}_{2\pi}
\{Z^\pi_{ij},Z^\pi_{jk}\} \{\tau_{ij},\tau_{jk}\} \, ,
\end{equation}
where
\begin{equation}
Z^\pi_{ij} = \frac{m_\pi r}{3} [Y(r)-T(r)] (S_{ij}+\sigma_{ij}) \, .
\end{equation}
The values of $A^{PW}_{2\pi}$ and $A_R$ are somewhat weaker than in UIX, but the energies of nuclei and saturation density of nuclear matter are virtually unchanged.

\section{Quantum Monte Carlo methods}
\label{sec:qmc}
A necessary step in computing the properties of atomic nuclei from the Hamiltonians discussed above is solving the many-body Schrödinger equation. In this work, we adopt continuum quantum Monte Carlo (QMC) techniques, a class of stochastic methods that provide an accurate and nonperturbative treatment of strongly interacting nuclear systems. A key advantage of QMC approaches in continuum coordinate space is their ability to handle realistic nuclear interactions featuring strong short-range repulsion and tensor components, making them particularly well suited for describing both the long-range structure and the short-range correlations of nuclear wave functions within a unified framework. Among the various QMC implementations, we focus here on a comparison between variational Monte Carlo (VMC) and auxiliary field diffusion Monte Carlo (AFDMC), which are discussed in more detail in this section. It is worth noting that Green's function Monte Carlo (GFMC) is also a member of the QMC family and has been used to obtain highly accurate ground-state energies, spectra, and electroweak observables for nuclei up to \(A \approx 12\text{--}13\)~\cite{Carlson:2014vla}. Although GFMC provides benchmark-quality results for light systems, we do not present calculations with this method here. Instead, we focus on VMC and AFDMC, noting that VMC already provides sufficient accuracy for spatial and momentum distributions~\cite{Wiringa:2013ala}.

\subsection{Variational Monte Carlo}
A VMC calculation finds an upper bound, $E_V$, to the ground-state energy, $E_0$, of the Hamiltonian by evaluating the expectation value of $H$ with a variational wave function, $\Psi_V$:
\begin{equation}
E_V = \frac{\langle \Psi_V | H | \Psi_V \rangle}
{\langle \Psi_V | \Psi_V \rangle} \geq E_0 \, .
\end{equation}

The optimal values of the variational parameters defining $\Psi_V$ are determined by minimizing $E_V$, and the corresponding minimum is taken as the approximate ground-state energy. The multidimensional integral over the $3A$ spatial coordinates, $R = \{\mathbf{r}_1,\dots,\mathbf{r}_A\}$, is evaluated using Metropolis--Hastings Monte Carlo techniques~\cite{Metropolis:1953}, from which the method derives its VMC designation. On the other hand, the summation over the exponentially many spin--isospin components is carried out explicitly. Specifically, the wave-function samples are generated, for each nucleus, by means of a random walk guided by 
\begin{equation}
\pi_V(R) = \frac{|\Psi_V(R)|^2} {\langle \Psi_V | \Psi_V\rangle}
\end{equation}
After an initial equilibration, a trial move is proposed in which each particle is randomly displaced within a box of side length $1.2-1.4$ fm. The probability distribution $\pi_V(R)$ is then evaluated for the proposed configuration and compared with its value for the previous one, and the move is accepted or rejected according to the Metropolis algorithm. After every ten attempted moves, the configuration is stored, including the $x$, $y$, and $z$ coordinates of each particle, with the center of mass set to zero, as well as the probability for each particle to be a neutron or a proton. With this choice of box size, the acceptance rate is approximately 50\%, and the number of generated configurations ranges from $10^6$ for $^3$He to $10^4$ for $^{12}$C.

A typical variational ansatz for light nuclei is~\cite{Wiringa:2000gb}
\begin{equation}
   |\Psi_V\rangle = {\cal S}\prod_{i<j}^A
      \left[1 + U_{ij} + \sum_{k\neq i,j}^{A} \tilde{U}_{ijk} \right]
                    |\Psi_J\rangle \, ,
\label{eq:psiv}
\end{equation}
where $U_{ij}$ and $U_{ijk}$ are noncommuting two- and three-body correlation operators induced by the dominant components of $v_{ij}$ and $V_{ijk}$, respectively, and ${\cal S}$ is a symmetrization operator. The Jastrow wave function is given by
\begin{equation}
   |\Psi_J\rangle = {\cal A}\prod_{i<j}f_c(r_{ij}) |\Phi_A(J^\pi;T T_z)\rangle \, .
\end{equation}
Here $\cal A$ is the antisymmetrization operator and the single-particle $A$-body wave function $\Phi_A(J^\pi;T T_z)$ is constructed from a sum of components with different spatial symmetries, all consistent with the quantum numbers of the state of interest. The product of central two-body correlation functions $f_c(r_{ij})$ over all pairs keeps nucleons apart, thereby accounting for the strong short-range repulsion of the interaction. For $p$-shell nuclei, there are in fact three distinct central pair-correlation functions, $f_{ss}$, $f_{sp}$, and $f_{pp}$, depending on whether both particles are in the $s$-shell core ($ss$), both are in the $p$-shell valence space ($pp$), or one belongs to each ($sp$). Their inclusion is essential for modeling the $\alpha$-clustering structure of $p$-shell nuclei. The long-range behavior of $f_c$, together with any single-particle radial dependence in $\Phi_A$---which, to ensure translational invariance, is written in terms of coordinates relative to the center of mass of the $s$-shell core---controls the finite spatial extent of the nucleus.  The antisymmetrization is carried out explicitly by summing over all partitions of the nucleons into s- and p-shell orbitals.

The structure of the two-body correlation operator reflects that of the $NN$ interaction:
\begin{equation}
   U_{ij} = \sum_{p=2,6} u_{p}(r_{ij}) O^{p}_{ij} \, ,
   \label{eq:U_2b}
\end{equation}
where the operators $O^{p}_{ij}$ correspond to the leading spin, isospin, spin-isospin, tensor, and tensor-isospin components of $v_{ij}$. The radial functions $f_c(r)$ and $u_p(r)$ are obtained by numerically solving a set of six Schr\"odinger-like equations: two single-channel equations for $S=0$, $T=0$ or 1, and two coupled-channel equations for $S=1$, $T=0$ or 1, the latter generating the important tensor correlations~\cite{Wiringa:1991kp}. These equations contain the bare $v_{ij}$ together with parametrized Lagrange multipliers used to impose long-range boundary conditions on the exponential decay and on the tensor-to-central ratios.

Perturbation theory is used to motivate the form of the three-body correlation operator,
\begin{equation}
\tilde{U}_{ijk} = -\epsilon \, \tilde{V}_{ijk}(\tilde{r}_{ij},
                     \tilde{r}_{jk}, \tilde{r}_{ki}) \, ,
\label{eq:U_3b}
\end{equation}
where $\tilde{r} = y r$, with $y$ a scaling parameter, and $\epsilon$ a (small, positive) strength parameter. The operator $\tilde{V}_{ijk}$ includes the dominant short-range repulsion and the anticommutator component of the two-pion-exchange contribution to the three-nucleon potential. Consequently, $\tilde{U}_{ijk}$ inherits the same spin, isospin, and tensor structure as $\tilde{V}_{ijk}$.

The variational parameters entering $f_{ss}$, $U_{ij}$, and $\tilde{U}_{ijk}$ are chosen to minimize the energy of the $s$-shell nucleus $^4$He. For $p$-shell nuclei from $^6$Li to $^{12}$C, these parameters are kept approximately constant, while the additional parameters entering $f_{sp}$, $f_{pp}$, and the single-particle radial dependence of $\Phi_A$ are optimized to minimize the energy of these systems, subject to the constraint that the proton and neutron root-mean-square radii remain close to those obtained from more accurate GFMC calculations~\cite{Pudliner:1997ck,Wiringa:2000gb}.

\subsection{Auxiliary field diffusion Monte Carlo}
The starting point of AFDMC calculations is a variational wave function similar to that of Eq.~\eqref{eq:psiv}, but with the correlation operator approximated as
\begin{align}
\hat{F} &=
\prod_{i<j}f_c(r_{ij})
\prod_{i<j<k}f_c(r_{ij},r_{ik},r_{jk}) \nonumber\\
& \times \Big[1 + \sum_{i<j}U_{ij} + \sum_{i<j<k}U_{ijk}\Big]
\label{eq:corr} \, ,
\end{align}
where the same notation $f_c$ is used for the scalar two- and three-body correlations. The explicit forms of $U_{ij}$ and $U_{ijk}$ are given in Eqs.~\eqref{eq:U_2b} and~\eqref{eq:U_3b}, respectively. To retain a computational cost that scales polynomially with the number of nucleons, following Ref.~\cite{Gandolfi:2014ewa}, the spin-isospin-dependent correlations are linearized with respect to Eq.~\eqref{eq:psiv}. In particular, products of correlation operators such as $U_{ij}U_{jk}$ are neglected, and the spin-isospin structure is kept only to linear order.

The AFDMC method projects out the lowest-energy state, $\Psi_0$, with the same quantum numbers as the variational ansatz according to
\begin{align}
|\Psi_0 \rangle \propto \lim_{\tau \to \infty} e^{-(H - E_V)\tau}|\Psi_V\rangle \, .
\end{align}
In this expression, $\tau$ denotes the imaginary time, while $E_V$ is an energy offset introduced to control the normalization and is typically chosen to be the variational energy. The direct computation of the propagator $e^{-H\tau}$ for arbitrary values of $\tau$ is not feasible for nuclear Hamiltonians. The imaginary-time evolution is therefore divided into a sequence of small time steps, $\delta\tau = \tau/N$, with $N$ large enough that the kinetic- and potential-energy operators entering $H$ can be exponentiated separately, thereby yielding a short-time propagator. More details on this procedure can be found in Refs.~\cite{Carlson:2014vla,Gandolfi:2020pbj}.

AFDMC keeps a computational cost that scales polynomially with the number of nucleons by representing the spin-isospin degrees of freedom as outer products of single-particle states. To preserve this structure during the imaginary-time propagation, Hubbard--Stratonovich transformations are employed to linearize the quadratic spin-isospin operators entering realistic nuclear potentials~\cite{Schmidt:1999lik}. While $NN$ interactions containing the first seven spin-isospin operators defined in Eq.~\eqref{eq:oper} can be treated in this way, the application of the Hubbard--Stratonovich transformation to the isospin-dependent spin-orbit term is presently not feasible. To circumvent this difficulty, we perform the imaginary-time propagation with a modified $NN$ interaction~\cite{Gnech:2024qru} in which
\begin{equation}
v^8(r_{ij}) \mathbf{L}\cdot\mathbf{S} \times \tau_{ij} \to 
\alpha\, v^8(r_{ij}) \mathbf{L}\cdot\mathbf{S}\,.
\label{eq:spin_orbit_prop}
\end{equation}
The parameter $\alpha$ is determined by requiring the expectation values of the original and modified AV8P interactions to be equal,
\begin{equation}
\langle v^8(r_{ij}) \mathbf{L}\cdot\mathbf{S} \times \tau_{ij} \rangle
=
\alpha \langle v^8(r_{ij}) \mathbf{L}\cdot\mathbf{S} \rangle \, .
\end{equation}

Analogously, the commutator term \(V_{ijk}^{\Delta,c}\) in Eq.~\eqref{eq:FM} contains cubic spin-isospin operators, which cannot at present be treated by continuous Hubbard--Stratonovich transformations. As in Ref.~\cite{Gnech:2024qru}, we follow the same strategy adopted for the isospin-dependent spin-orbit term of the \(NN\) interaction and perform the imaginary-time propagation with a modified three-nucleon potential,
\begin{equation}
V_{ijk}^{\Delta,c } \to \beta \, V_{ijk}^{\Delta,a} \, .
\label{eq:cubic_prop}
\end{equation}
The parameter \(\beta\) is again determined by requiring the expectation values of the original and modified \(V_{ijk}^{\Delta}\) terms to be equal,
\begin{equation}
\langle V_{ijk}^{\Delta,c } \rangle = \beta \langle V_{ijk}^{\Delta,a } \rangle \, .
\end{equation}

Finally, we note that within AFDMC, expectation values of operators that do not commute with the Hamiltonian---such as the spatial and momentum distributions, and the Wigner functions discussed in the next section---are estimated to first order in perturbation theory as
\begin{equation}
\frac{\langle\Psi(\tau) | O | \Psi(\tau)\rangle}{\langle\Psi(\tau) | \Psi(\tau)\rangle} \simeq 
2 \frac{\langle\Psi_V | O | \Psi(\tau)\rangle}{\langle\Psi_V | \Psi(\tau)\rangle} - \frac{\langle\Psi_V | O | \Psi_V\rangle}{\langle\Psi_V | \Psi_V\rangle}\, .
\label{eq:pc}
\end{equation}

\section{Wigner distributions}
\label{sec:wigner}

The single-nucleon off-diagonal density matrix encodes the most general one-body information accessible within a many-body framework. In coordinate space, it is defined as
\begin{align}
\rho_N(\mathbf{r}', \mathbf{r})
&= \sum_{i=1}^A \int dR \, d\mathbf{r}_i' \,
\Psi_0^\ast(\mathbf{r}_1, \ldots, \mathbf{r}_i', \ldots, \mathbf{r}_A)
\nonumber \\
& \times  \delta(\mathbf{r}' - \mathbf{r}_i') \delta(\mathbf{r} - \mathbf{r}_i)
P_N(i)\,
\Psi_0(R) \,.
\end{align}
where, for brevity, spin-isospin coordinates have been suppressed. The proton and neutron projection operators are given by
\begin{equation}
P_p(i) = \frac{1 + \tau_i^z}{2} \,, \qquad
P_n(i) = \frac{1 - \tau_i^z}{2} \,.
\label{eq:projectors}
\end{equation}

The diagonal component of the coordinate-space density is obtained as $\rho_N(\mathbf{r}) \equiv \rho_N(\mathbf{r}, \mathbf{r})$. When spin-orientation information is omitted, the density reduces to a scalar quantity. For a spherically symmetric ground state, this density depends solely on the radial coordinate $r = |\mathbf{r}|$. The corresponding radial density is then determined by averaging over the solid angle:
\begin{equation}
\rho_N(r) = \frac{1}{4\pi} \int d\Omega_r \, \rho_N(\mathbf{r}) \,.
\end{equation}

The single-nucleon momentum distribution is defined as the Fourier transform of the off-diagonal one-body density matrix:
\begin{equation}
n_N(\mathbf{k}) = \int d\mathbf{r} \, d\mathbf{r}' \, \rho_N(\mathbf{r}', \mathbf{r}) \, e^{-i\mathbf{k} \cdot (\mathbf{r} - \mathbf{r}')} \,.
\end{equation}

The Wigner quasi-probability distribution is defined as
\begin{equation}
W_N(\mathbf{r}, \mathbf{k}) = \int d\mathbf{s} \, e^{-i\mathbf{k} \cdot \mathbf{s}} \, \rho_N\left( \mathbf{r} + \frac{\mathbf{s}}{2}, \mathbf{r} - \frac{\mathbf{s}}{2} \right) .
\end{equation}
Since we will evaluate this quantity using QMC methods, it is convenient to express it directly in terms of the many-body wave function:
\begin{align}
&W_N(\mathbf{r}, \mathbf{k}) = \sum_{i=1}^A \int \! dR \, d\mathbf{s} \, e^{-i\mathbf{k} \cdot \mathbf{s}} \, \delta\left(\mathbf{r}_i - \mathbf{r} + \frac{\mathbf{s}}{2}\right) \nonumber\\
& \quad \times \Psi_0^\ast\left(\mathbf{r}_1, \ldots, \mathbf{r}_i + \mathbf{s}, \ldots, \mathbf{r}_A\right) P_N(i) \Psi_0(R) \,.
\label{eq:wigner_wave}
\end{align}

For systems with no preferred orientation, $W_N(\mathbf{r}, \mathbf{k})$ depends only on the magnitudes $r = |\mathbf{r}|$ and $k = |\mathbf{k}|$, and the cosine of the angle between them, $\cos\theta_{rk}$. If one is not interested in the angular correlations between $\mathbf{r}$ and $\mathbf{k}$, the Wigner function can be reduced to
\begin{align}
W_N(r,k) &= \frac{1}{8\pi^2} \int d\Omega_r \int d\Omega_k \, W_N(\mathbf{r}, \mathbf{k})
\nonumber \\
&= \int_{-1}^{1} d(\cos\theta_{rk}) \, W_N(r, k, \cos\theta_{rk}) \,,
\end{align}
which depends only on the magnitudes of the coordinate and the momentum. This expression implicitly averages over the spin degrees of freedom that are otherwise retained in the full matrix-valued Wigner function.

From Eq.~\eqref{eq:wigner_wave}, it is apparent that the QMC calculation of the Wigner distribution proceeds analogously to that of the momentum distribution, discussed extensively in Refs.~\cite{Wiringa:2013ala,Carlson:2014vla}. However, retaining coordinate-space information requires binning as a function of the distance
$|\mathbf{r}| = |\mathbf{r}_i + \mathbf{s}/2|$.
To reduce statistical noise, it is beneficial to displace the bra and ket ground-state wave functions symmetrically, each by half of the total displacement. Such a procedure, however, is only applicable at the VMC level and not in AFDMC. In the latter case, we therefore rely on a large number of Monte Carlo samples to reduce the statistical noise associated with the asymmetric estimator. To assess the effect of using the symmetric estimator, we have implemented this capability in the AFDMC code, although it is limited to the variational step.

The Wigner function is real, as a direct consequence of the Hermiticity of the one-body density matrix. However, it is not positive definite and therefore cannot be interpreted as a classical probability distribution in phase space. Negative regions arise from the intrinsically quantum-mechanical nature of the nuclear many-body wave function, reflecting interference effects associated with its nonlocal structure.

Despite this, the Wigner function has the important property that its marginal distributions reproduce physically meaningful observables. Integration over the spatial coordinate yields the momentum distribution:
\begin{equation}
\int d\mathbf{r} \, W_N(\mathbf{r}, \mathbf{k}) = n_N(\mathbf{k}) \,,
\end{equation}
while integration over momentum recovers the local density:
\begin{equation}
\int \frac{d\mathbf{k}}{(2\pi)^3} W_N(\mathbf{r}, \mathbf{k}) = \rho_N(\mathbf{r}) \,.
\end{equation}
It therefore provides a unified framework in which coordinate- and momentum-space information are simultaneously encoded. Furthermore, the Wigner function can be used to reconstruct the full off-diagonal density matrix:
\begin{equation}
\rho_N(\mathbf{r}', \mathbf{r}) = \int \frac{d\mathbf{k}}{(2\pi)^3} W_N\left(\frac{\mathbf{r} + \mathbf{r}'}{2}, \mathbf{k}\right) e^{i\mathbf{k} \cdot (\mathbf{r}' - \mathbf{r})} \,.
\end{equation}

In finite nuclei, the Wigner function captures both the long-range and short-range aspects of nuclear dynamics. Its dependence on the radial coordinate reflects the shell structure of the atomic nucleus and its spatial inhomogeneities~\cite{Nikolakopoulos:2025fgw}, while its high-momentum components encode short-range correlations generated by the underlying nuclear interaction. Being constructed from the off-diagonal density matrix, the Wigner function is particularly sensitive to nonlocal correlations beyond the independent-particle picture, offering a direct connection between spatial structure and the high-momentum behavior of the nuclear wave function.

\section{Gaussian Process Emulation}
\label{sec:GPs}
The direct QMC evaluation of the Wigner function for arbitrary values of $\mathbf{k}$ and $\mathbf{r}$ is too computationally expensive for in-line implementation in event-generator applications. However, such information is needed, for example, to simulate the propagation of the struck nucleon after the elementary lepton-nucleus scattering vertex, as in ACHILLES~\cite{Isaacson:2020wlx,Isaacson:2022cwh}. While angle-averaged Wigner functions can be interpolated using two-dimensional tables, retaining the full dependence on the relative orientation of $\mathbf{r}$ and $\mathbf{k}$ requires a higher-dimensional representation that cannot be handled in this way. Here, we opt for an \emph{emulation} of the Wigner function by training a Gaussian-process (GP) model on precomputed values of the function. The predictive mean of the trained GP model then provides an efficient
interpolant that can be called by the event-generator application.

A GP is formally defined as a collection of random variables, any finite subset of which has a joint normal distribution \cite{Rasmussen:2006}. Informally, a GP may be viewed as a normal distribution over functions, generalizing the notion of a finite-dimensional multivariate normal distribution to an infinite-dimensional setting. The usefulness of this construction can be understood from this perspective: just as, in a finite-dimensional normal distribution, specifying the values of some coordinates induces an updated normal distribution over the remaining coordinates conditional on those values, in a GP, observing possibly noisy function values at a finite set of points in the domain induces an updated, “trained” GP over function values at other points, conditioned on the observations. Moreover, the inherently stochastic nature of this prediction provides an uncertainty estimate that quantifies interpolation uncertainty and can be calibrated to yield probabilistically meaningful Bayesian credible regions \cite{Graziani_KMV2025}.

A GP model of a scalar function over a domain $\Omega$ is specified by a mean function $\mu(\cdot):\Omega\to\mathbb{R}$ and a covariance function $C(\cdot,\cdot):\Omega\times\Omega\to\mathbb{R}$. The latter is the analog of the covariance matrix in finite-dimensional normal theory and must be positive semidefinite as an integral kernel, although it may take negative pointwise values. Trained predictive distributions are constructed by evaluating these functions at the observed data points and using the resulting vectors and matrices to obtain the updated mean and covariance of the conditioned GP according to the formulas in \cite[][Chapter 2]{Rasmussen:2006}.

The technical challenge in constructing a GP emulator of the Wigner function is that the function must satisfy a specific normalization condition,
\begin{equation}
\int \frac{d\bm{k}}{(2\pi)^3} d\bm{r} W_N(\bm{r},\bm{k})=A_N,
\end{equation}
where $A_N$ is the number of nucleons of type $N$. It is important for the GP interpolant to enforce this condition to high accuracy, since, when sampling from the distribution, one does not want the nucleon number to fluctuate. This normalization imposes a linear constraint on the GP. Since GPs are closed under linear constraints \cite{jidling2017linearly}, imposing such a constraint on a primitive GP yields another GP. Constrained GP methods are surveyed in \cite{swiler2020survey}. To our knowledge, integral normalization constraints of this type have not been previously considered in this context. We provide an approach to this constraint problem here.

Consider the one-dimensional spherical case for the momentum distribution. The constraint is
\begin{equation}
I(n_N)\equiv\int_{0}^{\infty}dk\,k^{2}n_N(k)=A,\label{eq:Norm_Integral}
\end{equation}
where $A=2\pi^2A_N$. Given data for $n_N(k)$, one could fit an ordinary GP model to it; however, there would be no guarantee that functions sampled from the resulting GP would satisfy the normalization condition. Instead, suppose we start from a zero-mean GP with base covariance kernel
$\left\langle n_N(k_{1})n_N(k_{2})\right\rangle =C(k_{1},k_{2})$, with $k_1,k_2\in \mathbb{R}^+$. We can then update this kernel using a zero-noise “observation” of the normalization integral $I(n_N)$. The required covariance relations are
\begin{align}
\left\langle I\, n_N(k_{1})\right\rangle
&=
\int_{0}^{\infty}dk_{2}\,k_{2}^{2}
\left\langle n_N(k_{2})n_N(k_{1})\right\rangle \nonumber \\
&=
\int_{0}^{\infty}dk_{2}\,k_{2}^{2} C(k_{1},k_{2})
\equiv C_{1}(k_{1}),\label{eq:C_1}\\
\left\langle I^{2}\right\rangle
&=
\int_{0}^{\infty}dk_{1}\,k_{1}^{2}
\int_{0}^{\infty}dk_{2}\,k_{2}^{2}
\left\langle n_N(k_{2})n_N(k_{1})\right\rangle \nonumber \\
&=
\int_{0}^{\infty}dk_{1}
\int_{0}^{\infty}dk_{2}\,
k_{1}^{2}k_{2}^{2}C(k_{1},k_{2})
\equiv C_{0}.\label{eq:C_2}
\end{align}
Given a choice of $C(\cdot,\cdot)$ for which these integrals can be computed, the updated GP conditioned on the observation $I[n_N(\cdot)]=A$ has a nonzero mean function $\mu(k)$ given by the standard formula
\begin{equation}
\mu(k)=C_{1}(k) C_{0}^{-1} A,\label{eq:mu}
\end{equation}
and a covariance kernel $K(k_{1},k_{2})$ given by the equally standard formula
\begin{equation}
K(k_{1},k_{2})=
C(k_{1},k_{2})-C_{1}(k_{1})C_{0}^{-1}C_{1}(k_{2}).
\label{eq:K}
\end{equation}

Note that
\begin{equation}
\int_{0}^{\infty}dk\,k^2\mu(k)=C_{0}C_{0}^{-1}A=A,
\end{equation}
and functions $n_N(k)$ sampled from $GP\left[\mu(\cdot),K(\cdot,\cdot)\right]$ have normalizations $I[n_N(\cdot)]$ whose variance satisfies
\begin{align}
\langle \left(I-A\right)^2 \rangle &=
\int_{0}^{\infty}dk_{1}\int_{0}^{\infty}dk_{2}\,
k_{1}^{2}k_{2}^{2}K(k_{1},k_{2})\nonumber\\
&= C_{0}-C_{0}=0.
\end{align}
Thus, $I$ is fixed at its mean value $A$ with zero uncertainty. In effect, Eq.~\eqref{eq:K} shows that $K(\cdot,\cdot)$ has a zero mode corresponding to the constrained linear functional, namely integration with respect to the measure $k^2 dk$.

For this construction to work, the integrals in Eqs.~\eqref{eq:C_1} and \eqref{eq:C_2} must exist. This requirement is awkward because it precludes the use of a stationary kernel, $C(k_{1},k_{2})=g(k_{1}-k_{2})$, for which the integral in Eq.~\eqref{eq:C_2} diverges. This divergence is expected: a stationary kernel effectively assumes that the statistical behavior of the function is the same throughout the domain, whereas any
function with a finite normalization integral in Eq.~\eqref{eq:Norm_Integral} must be integrable with respect to the measure $k^2 dk$ and therefore decay sufficiently rapidly as $k\rightarrow\infty$. Another requirement is that the kernel integrals $C_1(k)$ and $C_0$ should be computable in closed form, rather than by numerical quadrature, since a quadrature-based approach would not be numerically efficient.

We have identified three valid (i.e., symmetric positive-semidefinite) kernels $C(\cdot,\cdot)$ with different asymptotic properties that satisfy these requirements. The first is a tapered squared-exponential kernel,
\begin{equation}
C^{\mathrm{TSE}}(k_1,k_2)=\beta\exp\left[-\frac{(k_1-k_2)^2}{\sigma^2}-\frac{k_1^2+k_2^2}{\gamma^2}\right],
\end{equation}
with $\beta>0$ and
\begin{equation}
a\equiv \sigma^{-2}+\gamma^{-2},\quad b\equiv-\sigma^{-2}.
\end{equation}
The corresponding kernel integrals are
\begin{align}
C^{\mathrm{TSE}}_1(k)&=\beta\frac{\sqrt{\pi}}{4}a^{-3/2}\left(1+\frac{2b^{2}k^{2}}{a}\right)\nonumber\\
&\quad\times\exp\left[-a\left(1-\frac{b^{2}}{a^{2}}\right)k^{2}\right]\mathrm{erfc}\left(a^{-1/2}bk\right)\nonumber\\
&\quad-\frac{bk}{2a^{2}}\exp\left(-ak^{2}\right),\\
C^{\mathrm{TSE}}_0&=\beta
\frac{1}{8}\left(a^{2}-b^{2}\right)^{-5/2}\left(a^{2}+2b^{2}\right)\nonumber\\
&\quad\times\left[\frac{\pi}{2}-\arcsin\left(b/a\right)\right]-\frac{3}{8}\left(a^{2}-b^{2}\right)^{-2}b.
\end{align}

The second is a rational kernel,
\begin{equation}
C^{\mathrm{R}}(k_1,k_2)=\beta(k_1+k_2+\alpha)^{-p},
\end{equation}
with $\beta>0$ and $p>6$, whose kernel integrals are
\begin{align}
C^{\mathrm{R}}_1(k)&=2\beta\frac{\Gamma(p-3)}{\Gamma(p)}(k+\alpha)^{-(p-3)},\\
C^{\mathrm{R}}_0&=4\beta\frac{\Gamma(p-6)}{\Gamma(p)}\alpha^{-(p-6)}.
\end{align}

The third is a variant rational kernel,
\begin{equation}
C^{\mathrm{R3}}(k_1,k_2)=\beta(k_1^3+k_2^3+\alpha)^{-p},
\end{equation}
with $\beta>0$ and $p>2$, whose kernel integrals are
\begin{align}
C^{\mathrm{R3}}_1(k)&=\frac{\beta}{3(p-1)}(k^3+\alpha)^{-(p-1)},\\
C^{\mathrm{R3}}_0&=\frac{\beta}{9(p-1)(p-2)}\alpha^{-(p-2)}.
\end{align}

In these expressions, the quantities $\beta$, $\sigma$, $\gamma$, $\alpha$, and $p$ are hyperparameters that may be varied to improve the fit to training data by maximizing the marginal likelihood \cite[][Chapter~2]{Rasmussen:2006}. This dependence may be indicated explicitly by writing, for example, $C(k_1,k_2;\theta)$, where $\theta$ denotes the set of hyperparameters.

A valid kernel over the spherically averaged phase space $(R,k)$ may be obtained straightforwardly through a tensor-product construction, for example
$C(R_1,k_1;R_2,k_2)=C^{R}(R_1,R_2)C^{TSE}(k_1,k_2)$.
In addition, positive linear combinations of valid kernels,
$C(\cdot,\cdot;\theta)\equiv\sum_{l}q_l C^l(\cdot,\cdot;\theta_l)$, with $q_l>0$, also yield valid GP kernels. In this way, one can construct models with sufficient flexibility to fit the training data.

We have implemented this GP scheme in \textit{Python} using the open-source \textit{GPyTorch} library \cite{GPyTorch}, which provides fast, accurate GP solves and GPU acceleration \cite{gardner2018gpytorch}.

\section{Results}
\label{sec:results}

\begin{figure*}[!htb]
\centering
\includegraphics[clip,width=\textwidth]{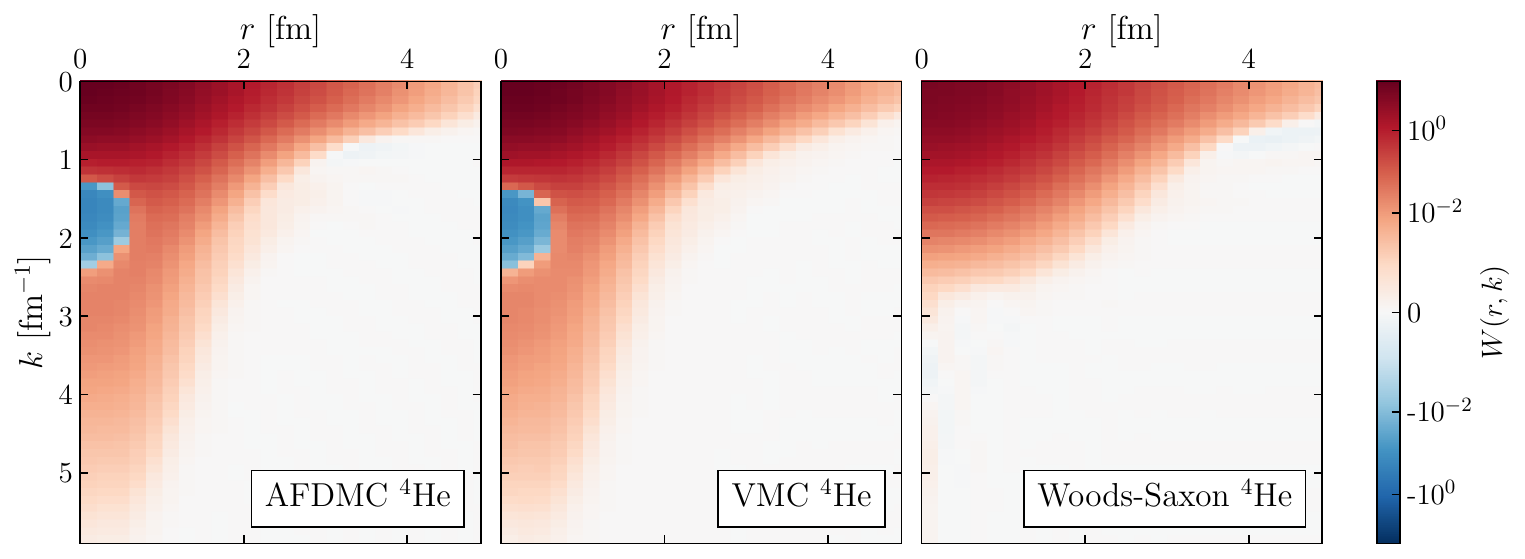}
\caption{%
Wigner distributions of spin-up protons in $^4$He obtained from AFDMC, VMC, and mean-field Woods-Saxon calculations, shown from left to right. The diverging color scale indicates positive (red) and negative (blue) values. The intensity reflects the magnitude of $W(r,k)$ using a symmetric logarithmic normalization, which enhances small-amplitude structures while preserving the sign.}
\label{fig:He4_wigner_comparison}
\end{figure*}

The Wigner function provides a phase-space representation of the nuclear wave function. It therefore makes it possible to identify how distinct dynamical mechanisms manifest themselves in specific regions of $(r,k)$. For instance, strength at small $r$ and large $k$ is associated with short-range correlations generated by the repulsive core and tensor components of the interaction, whereas the behavior of the quasi-probability distribution at larger radii and lower momenta is largely determined by long-range correlations and shell structure. In this section, we examine these features systematically and compare their behavior across different nuclei and interaction models.

\begin{figure}[!b]
\centering
\includegraphics[width=0.99\columnwidth]{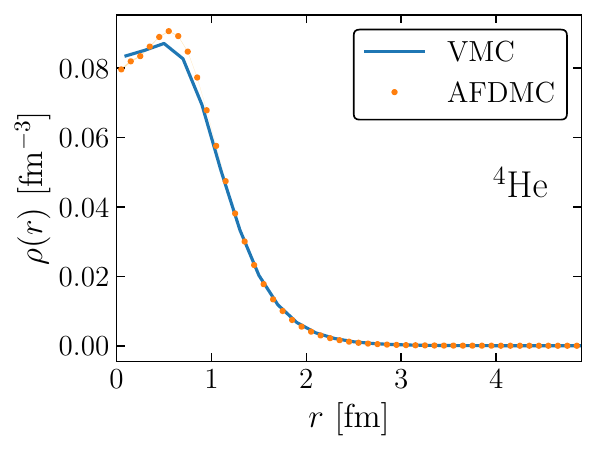}
\caption{%
Spatial density distribution of spin-up protons in $^4$He obtained from VMC and AFDMC calculations.
The VMC result is shown by the blue solid line, while the AFDMC result is shown by the orange points.
}
\label{fig:He4_densities}
\end{figure}

\begin{figure}[!b]
\centering
\includegraphics[width=0.99\columnwidth]{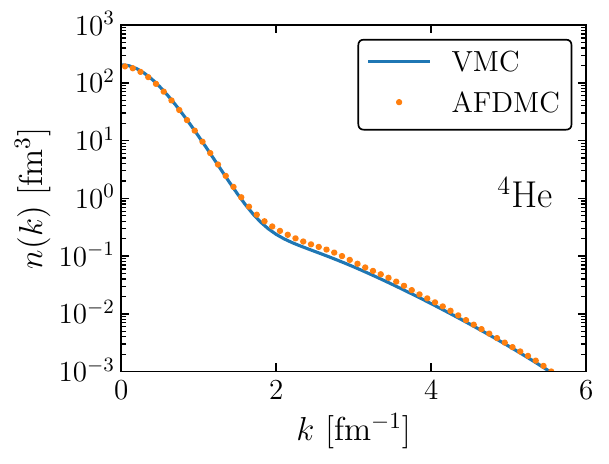}
\caption{%
Momentum distribution of spin-up protons in $^4$He obtained from the VMC and AFDMC calculations.
As in Fig.~\ref{fig:He4_densities}, the VMC result is shown by the blue solid line, while the AFDMC result is shown by the orange points.
}
\label{fig:He4_momentum_distributions}
\end{figure}

\begin{figure*}[!htb]
    \centering
    \includegraphics[clip,width=\textwidth]{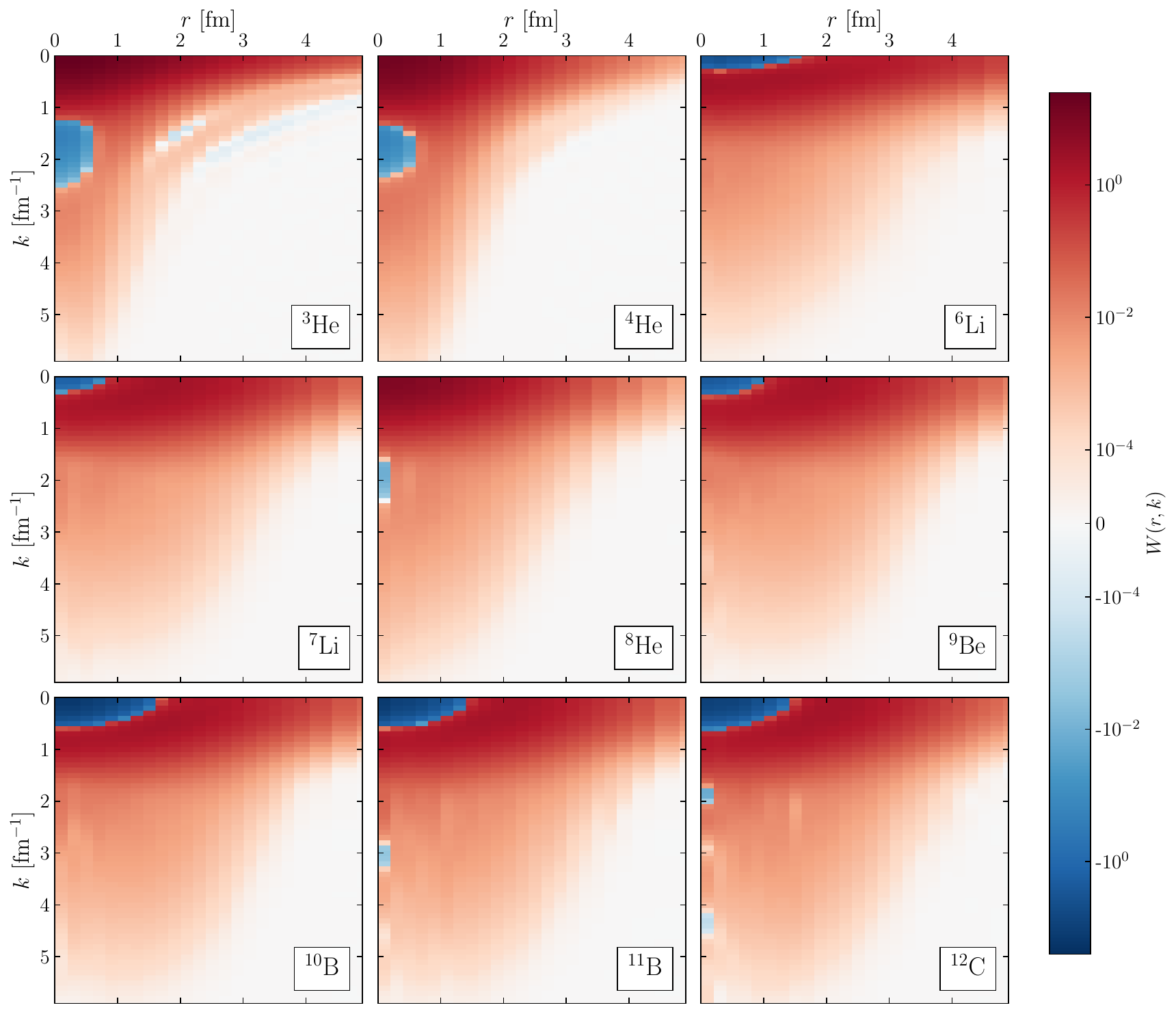}
    \caption{%
    VMC Wigner distributions of spin-up protons for selected nuclei with up to $A=12$ nucleons.
    }
    \label{fig:vmc_wigner_distributions}
\end{figure*}

In Fig.~\ref{fig:He4_wigner_comparison}, we compare the Wigner function of $^4$He obtained from AFDMC and VMC calculations with mean-field results based on a Woods-Saxon potential, obtained by setting $\hat{F}=1$ in Eq.~\eqref{eq:corr}. In the AFDMC case, the Wigner distribution is computed perturbatively using the estimator defined in Eq.~\eqref{eq:pc}. The AFDMC and VMC methods yield a consistent picture of the $^4$He nucleus. The high-momentum components of the wave function are mostly concentrated at distances smaller than $1\,\mathrm{fm}$, corresponding to the central region of the nucleus. This behavior is in stark contrast with the Woods-Saxon results, in which such high-momentum components are absent.

Both AFDMC and VMC calculations exhibit a negative region around $k \approx 2\,\mathrm{fm}^{-1}$ and small $r$. This feature is absent in the Woods-Saxon results, indicating that it is generated by dynamical correlations in the many-body wave function rather than by the independent-particle shell structure alone. The persistence of this negative region in calculations with simplified Argonne interactions, including Argonne $v_6^\prime$, Argonne $v_4^\prime$, and the spin-isospin-independent Argonne $v_1^\prime$~\cite{Wiringa:2002ja}, suggests that it should not be attributed specifically to spin-isospin-dependent correlations. Instead, we interpret it more generally as a correlation-induced interference effect associated with the short-range central correlations generated by the Argonne repulsive core.

\begin{figure*}[!htb]
    \centering
    \includegraphics[clip,width=0.75\textwidth]{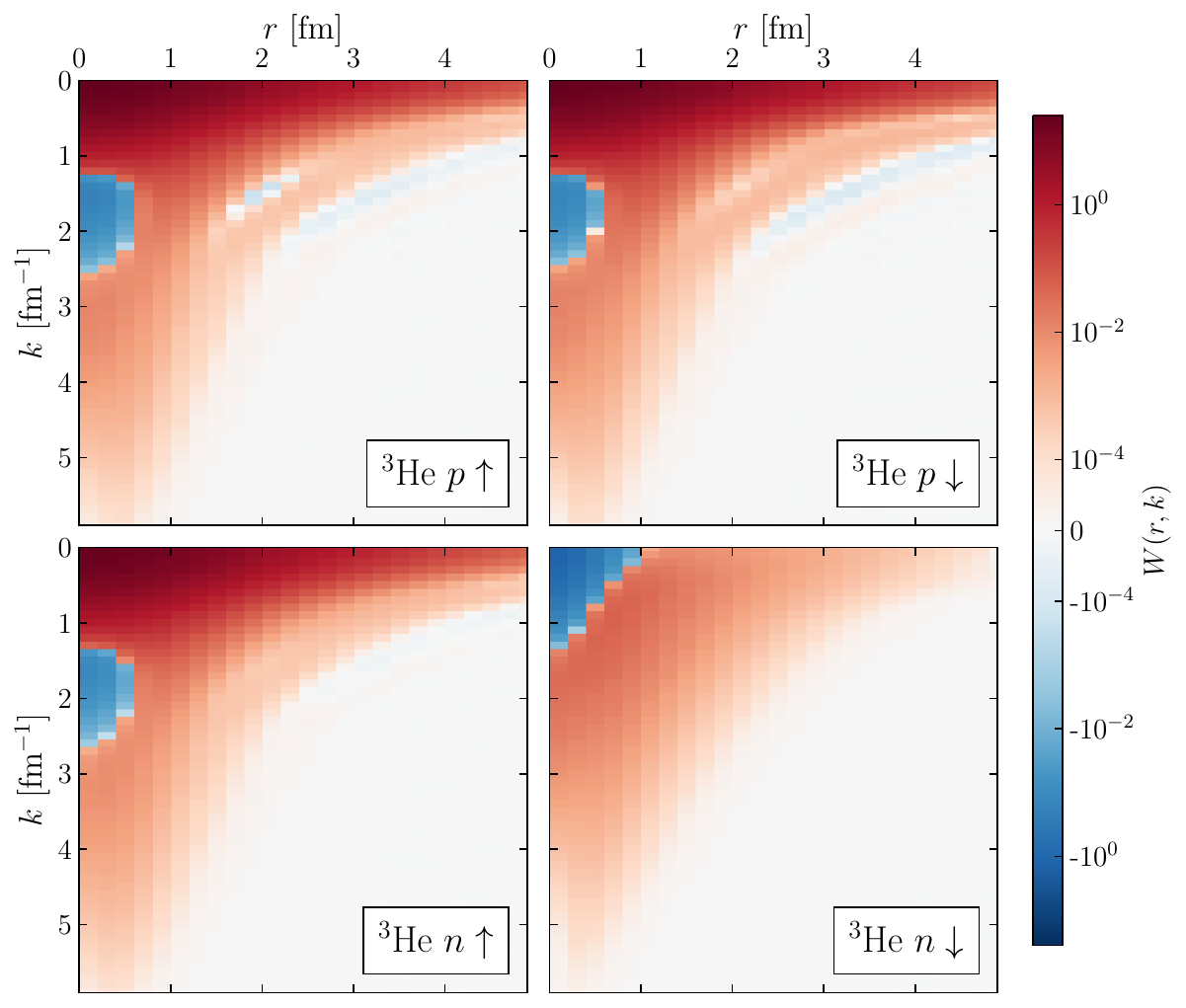}
    \caption{%
    Wigner distributions of all nucleons and spin orientations in $^3$He calculated from VMC.
    }
    \label{fig:He3_vmc}
\end{figure*}

The presence and redistribution of negative regions in the Wigner function leave signatures in the corresponding density and momentum distributions, highlighting the connection between phase-space structure and observable coordinate- and momentum-space quantities. Figure~\ref{fig:He4_densities} shows the single-nucleon spatial density distribution of $^4$He obtained with the VMC and AFDMC methods. While the overall agreement is very good, appreciable differences are observed in the radial region $r \lesssim 1\,\mathrm{fm}$, where the density is most sensitive to the short-distance structure of the wave function. In both the VMC and AFDMC results, the negative region of the Wigner function at small $r$ leads, after integration over momentum, to a corresponding depletion of the single-nucleon density. Using the same color scheme, Fig.~\ref{fig:He4_momentum_distributions} displays the corresponding momentum distributions. Again, the VMC and AFDMC results are broadly consistent, as expected given the similarity of the input Hamiltonians and the fact that both calculations describe the same compact $^4$He ground state. In particular, both approaches generate sizable high-momentum components, as is typical when using high-resolution $NN$ interactions of the Argonne family. The phase-space analysis provided by the Wigner function shows that these components originate primarily from regions close to the center of the nucleus.

VMC calculations of $W(k,r)$ for spin-up protons in selected nuclei with up to $A=12$ nucleons are illustrated in Fig.~\ref{fig:vmc_wigner_distributions}. For all nuclei, the projection $M$ of the total angular momentum $J$ along the $z$ axis is chosen to be maximal, $M=J$. The $s$-shell nuclei $^3$He and $^4$He display remarkably similar behavior, with the high-momentum components concentrated in spatial regions close to the center of mass of the nucleus. In contrast, for nuclei with $A\ge6$, the Wigner distribution extends over a much broader range of $r$ values, as nucleons begin to fill the $p$ shell. In addition, a negative region appears for all these nuclei at very small $k$, extending out to about $1\,\mathrm{fm}$ in $r$. The origin of this negative region has been traced by the authors of Ref.~\cite{Nikolakopoulos:2025fgw} to the angular-momentum structure of the occupied shell-model orbitals. In particular, the limiting behavior of the Wigner distribution at small $r$ and small $k$ implies that each shell contributes with a sign $(-1)^l$, where $l$ is the orbital angular momentum. As a consequence, the positive contribution from the $s$ shell is locally outweighed by the negative contribution from the occupied $p$-shell states, giving rise to the negative island near the origin of phase space. The absence of a pronounced negative island in the spin-up proton Wigner distribution of $^8$He is consistent with this shell-model interpretation. Although $^8$He is a $p$-shell nucleus, the additional nucleons beyond the $^4$He core are neutrons. The proton distribution therefore remains dominated by the $s_{1/2}$ shell, so the negative contribution associated with occupied proton $p$-shell orbitals is strongly suppressed.

For nuclei with $J=0$, the spin-up and spin-down nucleon distributions are identical. Similarly, for $T=0$ nuclei, proton and neutron distributions are the same, provided isospin is treated as a good quantum number and small charge-symmetry-breaking and charge-dependent terms are neglected in the correlation operator. An example in which neither condition holds is shown in Fig.~\ref{fig:He3_vmc} for $^3$He, where the Wigner distributions for all spin-isospin projections are compared. The Wigner distributions in $^3$He are generally similar to those of $^4$He, except for the spin-down neutron channel. This difference reflects our choice of the maximally polarized state, $M=J=+1/2$. In a simple $S$-wave picture, the two protons predominantly form a spin-zero pair, while the nuclear spin is carried by the spin-up neutron. As a result, the spin-down neutron distribution would vanish in the absence of spin-dependent correlations. Nonzero strength in this channel is therefore generated primarily by noncentral correlations, in particular tensor correlations, which admix higher-partial-wave components into the wave function. Consistently, the low-momentum components of the spin-down neutron momentum distribution are strongly suppressed~\cite{wigner_url}, as is also apparent from comparison with the spin-up neutron Wigner distribution in Fig.~\ref{fig:He3_vmc}. In the Wigner distribution, this suppression manifests itself through a pronounced negative component near the center of the nucleus at low momentum, which compensates the positive strength at larger distances. The corresponding momentum distributions obtained by integration over $r$ are illustrated in Fig.~1 of Ref.~\cite{Piarulli:2022ulk} where the shift from s-shell to p-shell nuclei is also quite noticeable.

\begin{figure*}[tbh]
\centering
\includegraphics[width=0.99\textwidth]{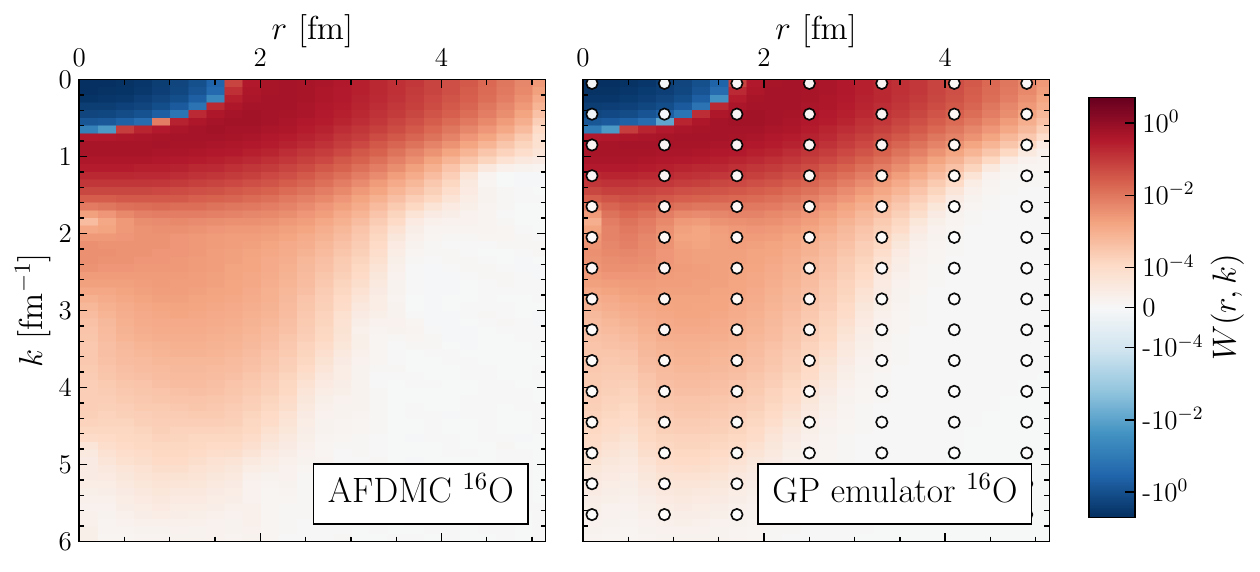}
\caption{
Wigner distribution of $^{16}$O obtained within AFDMC.
The left panel shows the original AFDMC result, while the right panel shows the corresponding GP-emulated distribution.
The filled circles indicate the coordinates used to train the GP emulator.
The color scheme is the same as in Fig.~\ref{fig:He4_wigner_comparison}.
}
\label{fig:O16_momentum_distributions}
\end{figure*}

The left panel of Fig.~\ref{fig:O16_momentum_distributions} shows the Wigner distribution of $^{16}$O obtained within the AFDMC framework. The case of $^{16}$O is particularly relevant for water-Cherenkov neutrino-oscillation experiments, such as Super-Kamiokande, T2K, and Hyper-Kamiokande, where neutrino interactions with oxygen constitute an essential component of the detected event sample. Calculations of this nucleus require many-body methods whose computational cost scales polynomially with the number of particles. In this respect, AFDMC enables a quantitative exploration of phase-space structure in nuclei beyond the reach of approaches based on an explicit summation over spin-isospin degrees of freedom.

The Wigner distribution of $^{16}$O displays a pronounced depletion at small $r$ and low momenta. This behavior, again interpreted as the negative contribution associated with occupied $p$-shell orbitals, is consistent with the VMC calculations of $p$-shell nuclei shown in Fig.~\ref{fig:vmc_wigner_distributions} and with the mean-field calculations of Ref.~\cite{Nikolakopoulos:2025fgw}. The strength localized at small $r$ and large $k$ reflects short-range dynamics induced by the repulsive core and tensor components of the nuclear interaction. At the same time, the strength extending over larger radii and finite momenta reflects the combined effects of shell structure and many-body correlations. In contrast to $^4$He, and more similarly to $^{12}$C, the high-momentum components are not confined as sharply to the center of the nucleus. This broader spatial distribution is consistent with the emergence of substructure, such as alpha-clustering correlations, although such correlations cannot be directly resolved from the angle-averaged Wigner distribution alone. The appearance of these characteristic phase-space patterns in $^{16}$O further demonstrates that the qualitative signatures identified in lighter nuclei persist in medium-mass systems.

The right panel of Fig.~\ref{fig:O16_momentum_distributions} displays the Wigner distribution interpolated using the GP emulator described in Sec.~\ref{sec:GPs}. The coordinates used to train the GP are shown as filled circles and correspond to one fourth of the points in the initial grid. The emulated Wigner distribution exhibits behavior remarkably close to the input distribution over the entire radial and momentum domain. In particular, the GP accurately reproduces both the negative region at small $r$ and small $k$, and the strength localized at small $r$ and large $k$ associated with short-range dynamics induced by the repulsive core and tensor components of the nuclear interaction. Importantly, the emulated Wigner distribution preserves the normalization by construction and is therefore readily employable in neutrino event generators.

 Tabulated values of $W(k,r)$ from the VMC calculations are available online~\cite{wigner_url}.

\section{Conclusions}
\label{sec:conclusions}

In this work, we have presented {\it ab initio} calculations of the Wigner quasi-probability distributions of selected light nuclei, up to $A=16$ nucleons. Compared with earlier studies~\cite{Prakash:1981ilg,Bonasera:1993zz,Cosyn:2021ber,Nikolakopoulos:2025fgw}, which were limited to mean-field approximations or included only short-range correlations, our calculations fully incorporate dynamical many-body correlations by employing continuum quantum Monte Carlo methods to solve the Schr\"dinger equation with high accuracy. We use high-resolution $NN$ potentials of the Argonne family, supplemented by consistent $3N$ forces, which are known to generate sizable high-momentum tails in nuclear momentum distributions~\cite{Wiringa:2013ala}.

The impact of short-range correlations is already apparent in the lightest nuclei considered here, $^3$He and $^4$He, whose Wigner distributions display remarkably similar features. In both cases, the Wigner function exhibits significant strength at large momenta. In addition, the short-range central correlations induced by the repulsive core of the interaction lead to the appearance of a small negative region around $k \approx 2\,\mathrm{fm}^{-1}$ and small $r$. We verified that this feature is absent in mean-field calculations based on a Woods-Saxon potential, indicating that it originates from dynamical correlations in the many-body wave function.

The shell structure of $p$-shell nuclei is also clearly reflected in their Wigner distributions. Compared with $s$-shell systems, the strength extends to much larger radii. In addition, we observe a negative region at very small momenta and $r \lesssim 1\,\mathrm{fm}$, which can be ascribed to the angular-momentum structure of the occupied shell-model orbitals~\cite{Nikolakopoulos:2025fgw}. Consistently with this interpretation, this negative region is absent in the spin-up proton Wigner distribution of $^8$He, which is dominated by nucleons occupying the $s$ shell.

The Wigner distributions of $^{12}$C and $^{16}$O reveal a qualitatively different localization pattern for the high-momentum components. Unlike in $^4$He, these components are not concentrated solely at very small radii, but are largest around $r \simeq 1.5\,\mathrm{fm}$. This behavior is consistent with the emergence of an $\alpha$-clustered structure in these nuclei: the high-momentum components originate primarily near the centers of the constituent $\alpha$ clusters. Although such correlations cannot be directly resolved in the angle-averaged Wigner distribution, their imprint is visible in the radial dependence of the high-momentum strength. The Wigner function therefore provides a unified representation in which long-range spatial structure, $\alpha$ clustering, shell effects, and short-range correlations can be simultaneously identified and analyzed.

In addition to their intrinsic interest, Wigner functions are directly relevant for neutrino event generators, where an accurate description of the propagation of nucleons through the intranuclear medium after the primary interaction vertex is essential~\cite{Isaacson:2020wlx,Isaacson:2022cwh}. They provide information on the joint distribution of finding a nucleon with momentum $|\mathbf{k}|$ at a distance $|\mathbf{r}|$ from the nuclear center of mass, thereby correlating the position and momentum sampled in the initial nuclear state.

Beyond neutrino event generators, nuclear Wigner distributions also provide a natural bridge between ab initio nuclear structure and descriptions of partonic degrees of freedom in finite nuclei. Recent work has used finite-nucleus Wigner distributions as an analogue of nuclear-matter occupation numbers to construct quark phase-space distributions through a convolution with the quark momentum distribution inside the nucleon~\cite{Nikolakopoulos:2025fgw}. This framework was used to investigate whether the low-momentum nucleon suppression associated with quarkyonic-matter scenarios could persist in large finite nuclei. The ab initio Wigner distributions presented here, which include dynamical many-body correlations beyond mean-field descriptions, may therefore provide useful microscopic input for future studies connecting nuclear structure with partonic degrees of freedom.

To facilitate their inclusion in neutrino event generators, and more generally in applications requiring phase-space nuclear structure information, we have developed a Gaussian-process emulator that provides a compact and efficient representation of the Wigner distributions. The accuracy of this representation is enhanced by imposing the exact normalization of the Wigner function. For the spherical nuclei discussed in this work, the angle-averaged distributions can also be stored and interpolated using two-dimensional tables, which are made available in an online repository~\cite{wigner_database}. However, retaining the full dependence on the relative orientation of $\mathbf{r}$ and $\mathbf{k}$ requires a higher-dimensional representation, for which direct tabulation becomes impractical. In this case, the Gaussian-process emulator provides a natural and flexible alternative.

\section{Acknowledgments}
We are grateful to Alexis Nikolakopoulos for helping us identify a bug in the AFDMC calculations that affected the Wigner distributions in an earlier version of this work.  We also thank Jane Kim for assistance in preparing the online tabulation. The present research is supported by the U.S. Department of Energy, Office of Science, Office of Nuclear Physics, under contracts DE-AC02-06CH11357 (A.~T., A.~L., R.~B.~W.), by the DOE Early Career Research Program (A. L.), by the Fermi Research Alliance, LLC under Contract No. DE-AC02-07CH11359 with the U.S. Department of Energy, Office of Science, Office of High Energy Physics (N.R.), by the NUCLEI SciDAC-5 project (A.~L, R.~B.~W.) and NeuCol programs (A.~L., N.~R.), and by grant PID2023-147458NB-C21 (A.~L.) funded by MCIN/AEI/10.13039/501100011033, and by the European Union. This research used resources of the Laboratory Computing Resource Center of Argonne National Laboratory.
N.R. is also supported by an ERC STG grant (“NuQNET”, grant No.101164195).
\bibliography{biblio}

\end{document}